\documentclass{PoS}

\title{Measurement of hard double-parton interactions with the ATLAS detector}

\ShortTitle{Measurement of hard double-parton interactions with the ATLAS detector}

\author{\speaker{Miroslav MYSKA} \textmd{ on behalf of the ATLAS Collaboration}\\
	Institute of Physics, Academy of Sciences of the Czech Republic,\\
	Na Slovance 2, 18221 Prague, Czech Republic\\
        E-mail: \email{myskam@fzu.cz}}

\abstract{The production of $W$ bosons in association with two jets in proton-proton collisions at the LHC with a centre-of-mass
energy of $\sqrt{s}$ = 7 TeV has been investigated using data collected with the ATLAS detector. The fraction of
events arising from double-parton scattering
has been measured through the transverse momentum balance between the two jets. The measurement provides new
information on the scaling of the rate of double-parton scattering processes
and constraints on proton transverse profile modeling.}

\FullConference{XXI International Workshop on Deep-Inelastic Scattering and Related Subjects\\
                 22-26 April, 2013\\
                 Marseilles, France}

\begin{document}

\section{Introduction}

The inelastic scattering of hadrons is usually modeled as a Single-Parton Interaction (SPI), 
i.e. a perturbatively calculated interaction of a pair of partons, each coming from a different hadron.
The remaining partons then result in a large amount of final state particles, generally called as an underlying event.
The possibility of having additional interactions among these remaining partons has been 
studied since early days of parton model in 70's. 
The simplest case of Double-Parton Interaction (DPI), where at least two pairs of partons interact at hard scales,
has been measured by several experiments, which
quantified the scaling of its production cross section. 
The latest measurements were done at the Fermilab's Tevatron
$p\bar{p}$ collider \cite{Abe:1997:prd56_3811,Abazov:2010:prd81_052012} and reached very high accuracy.

All experiments assumed 
that the two interactions are independent at parton level and that the expression
for the inclusive DPI cross section can be fully factorized:
\begin{equation}
 \sigma_{\rm DPI} = \frac{1}{1+\delta_{\rm ab}}\frac{\sigma_{\rm a} \sigma_{\rm b}}{\sigma_{\rm eff}}.
 \label{myska:eq:DPI_Xsec}
\end{equation} 
\noindent Here, a general case for processes a and b is considered 
and therefore one has to introduce a combinatorial factor $\delta_{\rm ab}$ avoiding the double counting 
of indistinguishable processes. The scaling denominator $\sigma_{\rm eff}$ is a measurable quantity 
depending primarily only on the type of the hadron. At first, the naive expectation replaced 
$\sigma_{\rm eff}$ by the given inelastic cross section. On the contrary, the measurements
found this quantity to be roughly from 5 to 20 mb. This comparison favors the model, in which
the hadron consists of a \textit{hard} parton core surrounded by a wider cloud of \textit{soft} partons.
In addition to the transverse hadron profile modeling, also inter-parton correlations play a non-negligible role.
Therefore, the precise evaluation of this scaling factor is a very important task and it is the goal of the 
ATLAS measurement \cite{Aad:2013bjm} described in the following.

\section{DPI contribution to $W$ + 2-jet production}

\begin{figure}[h!]
\centering
\vspace{-0.8cm}
 \begin{tabular}{cc}
  \begin{minipage}{.4\hsize}
    \centering
     \includegraphics[width=3.5cm]{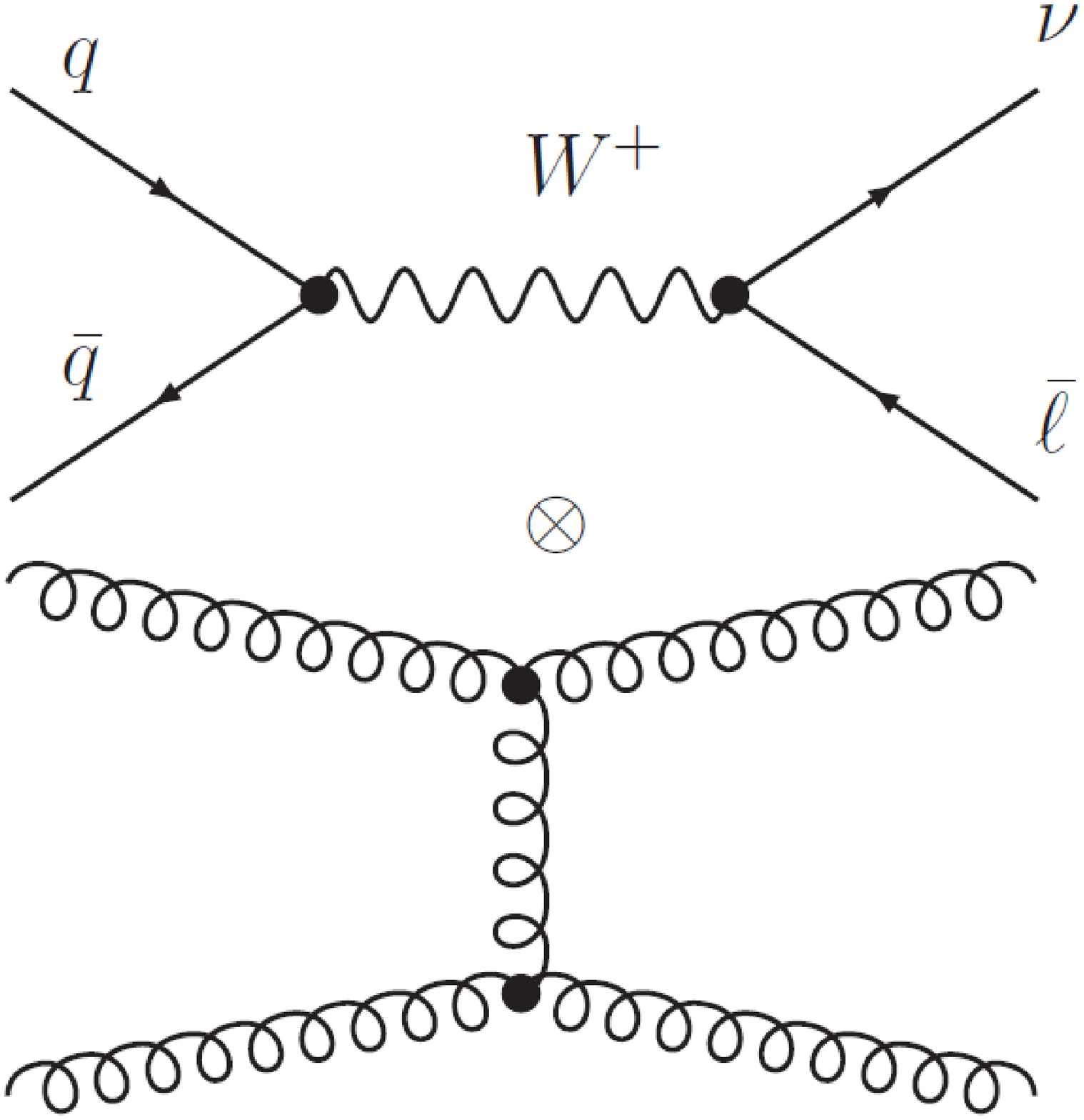}
  \end{minipage}

  \begin{minipage}{.4\hsize}
   \centering
     \includegraphics[width=3.5cm]{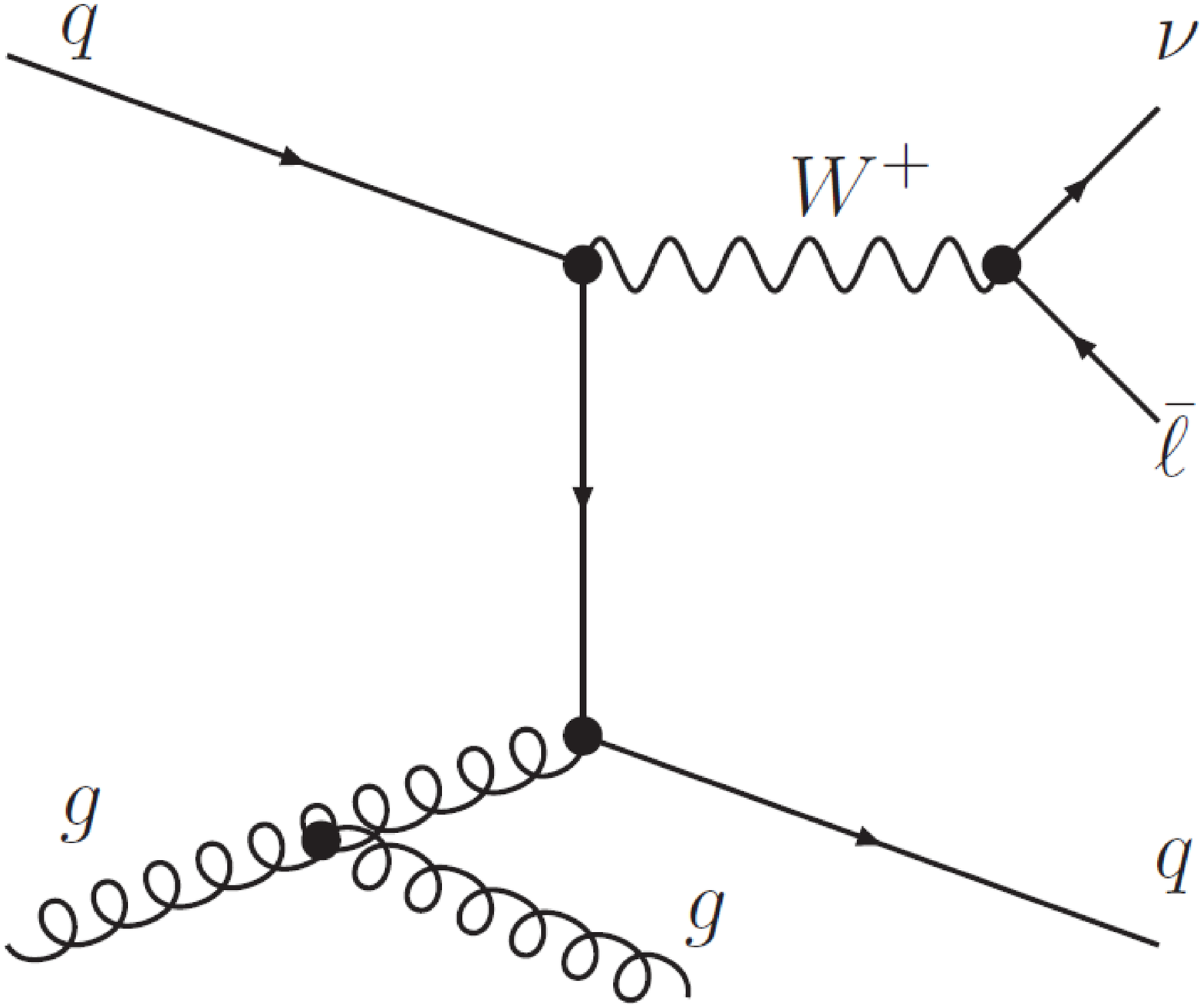}
   \end{minipage}
 \end{tabular}
\caption{Two main types of leading-order diagrams contributing to $W$ + 2-jet production \cite{Aad:2013bjm}. Left: DPI diagram consisting of independent $W$ + 0-jet production
combined with a QCD di-jet production. Right: SPI diagram containing two 
jet-initiating outgoing lines directly connected to the primary process.}
\label{myska:fig:SPI_DPI_diagrams}
\end{figure}

The specific DPI process investigated using the ATLAS detector \cite{Aad:2008zzm} is the production of a $W$ boson associated with exactly
two jets in proton-proton collision at a center-of-mass energy of 7 TeV. The gauge boson decays in either the electron or
muon channel. This study assumes the validity of the fully factorized model, represented here by Eq. (\ref{myska:eq:DPI_Xsec}),
and models the DPI event by combining two independent parton interactions. In the first one, as shown in the left plot in Fig. \ref{myska:fig:SPI_DPI_diagrams}, 
$W$ boson is produced in association with no jet in the studied phase space region, while the second interaction of type QCD 2$\rightarrow$2 produces exactly two jets.
The possibility of producing exactly one jet associated with the first interaction
and exactly one jet connected to the second interaction is highly suppressed and was found to be negligible.
Also the appropriate contribution from triple parton scattering is neglected.

However, the exclusive SPI process is still the leading production mechanism for this final state and phase space. In these events, outgoing quark or gluon lines 
initiating measured jets are directly associated with the primary interaction. An example of such a Feynman
diagram is shown in the right plot in Fig. \ref{myska:fig:SPI_DPI_diagrams}. 

\section{Event selection}

The analysis uses the entire 2010 ATLAS data, corresponding to the integrated luminosity
of 36 pb$^{-1}$. The selection of $W$ events is based on the reconstruction of isolated either electron,
required to have transverse momentum $p_{\rm T}$ $>$ 20 GeV and pseudorapidity $|\eta|$ $<$ 2.47, or muon,
satisfying $p_{\rm T}$ $>$ 20 GeV and $|\eta|$ $<$ 2.4.
Moreover, a missing transverse energy in the detector, $\vec{E}_{\rm T}^{\rm miss}$, is required to satisfy $|\vec{E}_{\rm T}^{\rm miss}|$ $>$ 25 GeV.
The reconstructed transverse mass of the $W$ boson, defined as
${m}_{\rm T} = [ 2 {p}_{\rm T}^{\rm l}|\vec{E}_{\rm T}^{\rm miss}| (1-\cos \Delta\phi_{\rm l,\vec{E}_{\rm T}^{\rm miss}}) ],$
has to exceed 40 GeV in order to suppress the unwanted background.

Jets were reconstructed using anti-$k_t$ algorithm with radius-like parameter R = 0.4 and were required to have
transverse momentum $p_{\rm T}$ $>$ 20 GeV and rapidity $|y|$ $<$ 2.8. 
In order to suppress jets stemming from additional proton collisions in the same bunch-crossing (pile-up),
the fraction of jet tracks associated with the primary vertex (JVF) is calculated for each jet located in the inner detector
and containing at least one track.
Jet is removed from the analysis if JVF $<$ 0.75. In addition, jets within $\Delta$R $<$ 0.5 
of the accepted lepton are also removed from the analysis.


\section{Method of $\sigma_{\rm eff}$ extraction}

The strategy for the evaluation of $\sigma_{\rm eff}$ is very straightforward. Rewriting of Eq. (\ref{myska:eq:DPI_Xsec})
to terms of $W$ + 2-jet final state leads to 
\begin{equation}
\sigma_{\rm eff} = \frac{\sigma_{W{\rm + 0j}}\sigma_{\rm 2j}}{\sigma_{W{\rm+2j}}^{\rm DPI}} 
= \frac{1}{{f}_{\rm DP}} \frac{{N}_{W{\rm +0j}}}{{N}_{W{\rm+2j}}}
\frac{{N}_{\rm 2j}}{\mathcal{L}_{\rm 2j}},
\label{myska:eq:sigma_eff_eval}
\end{equation} 
\noindent where cross sections are replaced by the numbers of events. Since the analysis assumes the fully factorized model of DPI,
all of the factors reflecting geometrical acceptance of the ATLAS detector and corrections for detector effects cancel in the ratio.
Also terms for luminosity and trigger efficiency corresponding to the $W$ + 2-jet and $W$ + 0-jet data samples cancel in the ratio. The remaining factor is
the integrated luminosity for the di-jet data sample, $\mathcal{L}_{\rm 2j}$.

While the integrated luminosity and the event numbers can be directly read from the appropriate data samples, where one requires 
events to contain exactly one reconstructed primary vertex, the factor $f_{\rm DP}$ has to be obtained by a further analysis.
It is defined as a fraction of DPI events with respect to the total number of exclusive SPI plus exclusive DPI events:
\begin{equation}
{f}_{\rm DP} = \frac{{N}_{W{\rm+2j}}^{\rm DPI}}{{N}_{W{\rm+2j}}} = \frac{{N}_{W{\rm+2j}}^{\rm DPI}}{{N}_{W{\rm+2j}}^{\rm SPI} + {N}_{W{\rm+2j}}^{\rm DPI}}.
\end{equation} 

\begin{figure}[h!]
\centering
 \begin{tabular}{cc}
  \begin{minipage}{.45\hsize}
    \centering
     \includegraphics[width=7cm]{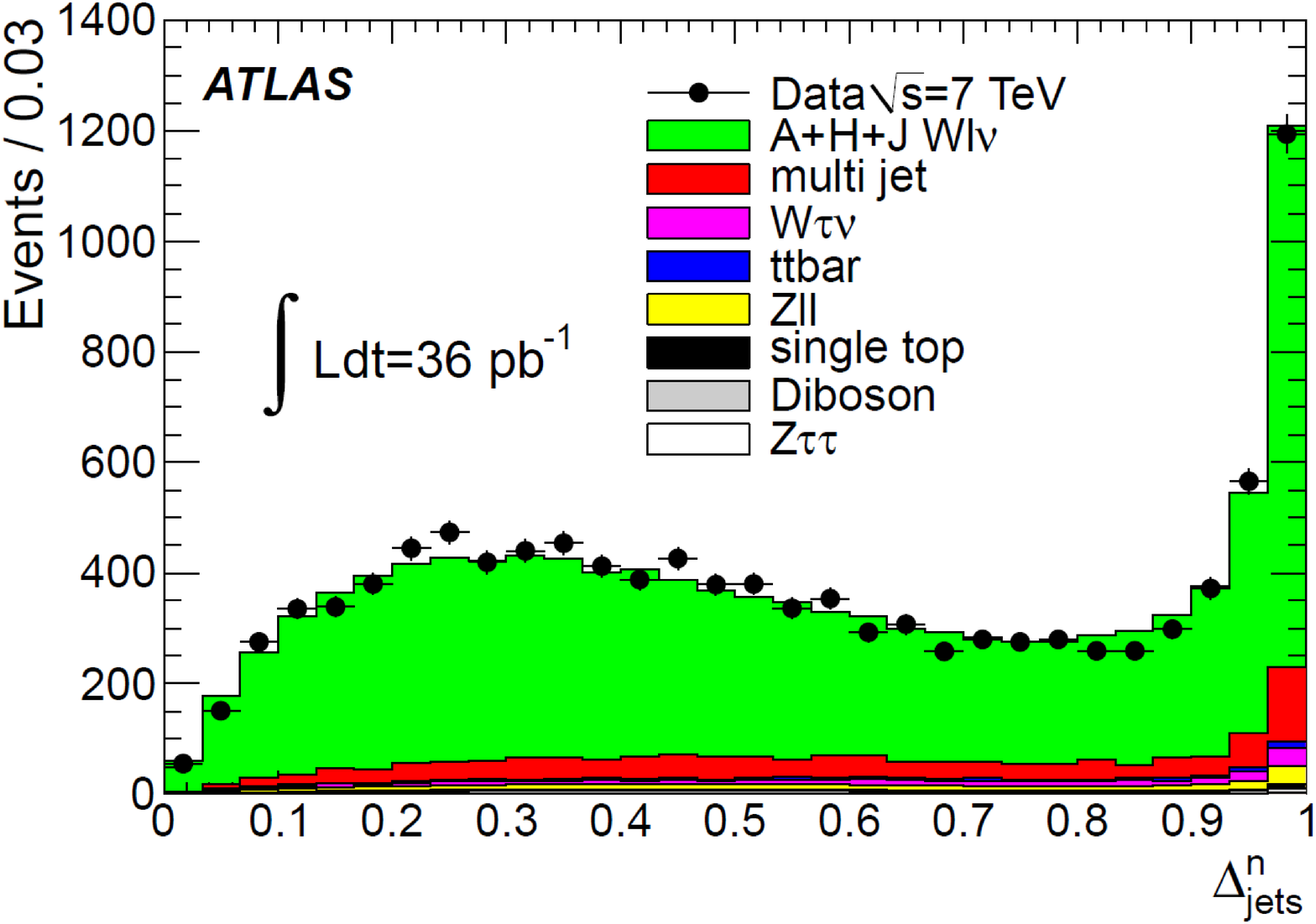}
  \end{minipage}

  \begin{minipage}{.5\hsize}
   \centering
     \includegraphics[width=7cm]{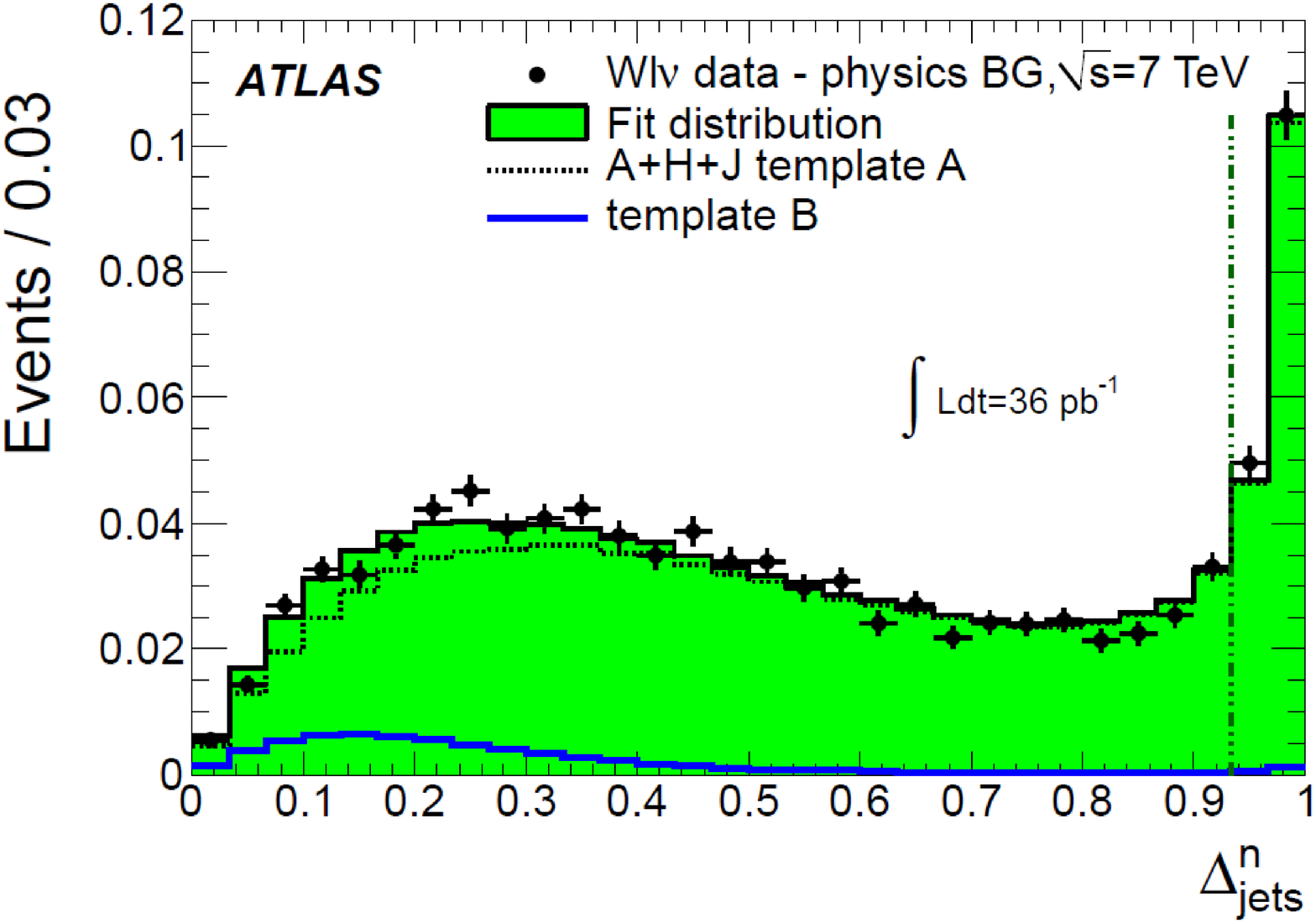}
   \end{minipage}
 \end{tabular}
\caption{$\Delta^{\rm n}_{\rm jets}$ distributions \cite{Aad:2013bjm}. Left: Comparison of
ATLAS data points with Monte Carlo simulated processes contributing to $W$ + 2-jet production. Right: Comparison of the  
inclusive SPI production (green area distribution), given by the best combination of exclusive SPI simulated (black dotted line)
and the di-jet ATLAS data (blue solid line) modeling DPI events, to background-subtracted ATLAS data points.}
\label{myska:fig:Imbalance}
\end{figure}

To extract the fraction of DPI events, two steps have to be performed.
First, the normalized jet-pair transverse momentum imbalance
defined as
\begin{equation}
\Delta^{\rm n}_{\rm jets} = \frac{|\vec{p}^{{\rm j}_1}_{\rm T} + \vec{p}^{{\rm j}_2}_{\rm T}|}
{|\vec{p}^{{\rm j}_1}_{\rm T}| + |\vec{p}^{{\rm j}_2}_{\rm T}|}
\end{equation} 
\noindent is chosen to be the most suitable quantity for discriminating between SPI and DPI events. Second, the event fraction coming from physics background processes 
is estimated using Monte Carlo simulations and the appropriate component of the $\Delta^{\rm n}_{\rm jets}$ distribution for all the ATLAS data is subtracted.
The left plot in Fig. \ref{myska:fig:Imbalance}
shows the $\Delta^{\rm n}_{\rm jets}$ distributions for all the ATLAS data and simulated processes contributing to the given final state.
The main component of the measured events is the inclusive SPI production containing both exclusive SPI and DPI events. The inclusive SPI
process contains at least one hard parton interaction, while the exclusive SPI or DPI events contain exactly the given number of scatters.
The background processes form around 19$\%$
of events in the electron channel and around 14$\%$ in the muon channel.

The quantity $f_{\rm DP}$ is obtained by comparing $\Delta^{\rm n}_{\rm jets}$ distribution for background-subtracted ATLAS data, 
corresponding to inclusive SPI $W$ + 2-jet production, with $\Delta^{\rm n}_{\rm jets}$ distributions for events modeling exclusive SPI (A) and exclusive DPI (B)
contributions. The two latter distributions are called templates and are both normalized to unity. Their linear combination
$(1-{f}_{\rm DP}) {\rm A} + {f}_{\rm DP} {\rm B}$
has to fit the normalized distribution for the ATLAS data. 
The template B is obtained using the di-jet ATLAS data, since the studied distribution is the same for di-jets coming from the same scatter, whether 
it is primary or secondary parton interaction. The correction for the jet-lepton isolation condition is found to be negligible.

The exclusive SPI event sample necessary for template A modeling is prepared using 
Alpgen generator \cite{Mangano:2003:jhep0307_001} interfaced with Herwig v6.510 \cite{Corcella:2001:jhep0101_010} 
and Jimmy v4.31 \cite{Butterworth:1996:zpc72_637} (A+H+J). 
The events are obtained by applying a parton-level cut
$p_{\rm T}^{\rm p}$ $<$ 15 GeV applied on all outgoing partons from any QCD interaction additional to the $W$ production.
This cut was chosen after a careful optimization. Its value was varied between 10 and 17.5 GeV, affecting the resulting fraction $f_{\rm DP}$ by around 5$\%$.
The complete rejection of all events with the additional hard parton interactions 
would inappropriately affect the studied $\Delta^{\rm n}_{\rm jets}$ distribution. The resulting jets would be much more
correlated than in the real case, where the underlying event smears their directions.

The right plot in Fig. \ref{myska:fig:Imbalance} shows the result of the best fit of the linear combination of prepared templates to the 
background-subtracted data. The two right-most bins, separated by a vertical line in the plot, were excluded from the fit,
since these events contain too much collinear jets.

\section{Result}

All the data used for the fit, whether real or simulated, may contain an arbitrary number of pile-up collisions. In order to avoid the dependence on the 
pile-up, a correction is evaluated using inclusive as well as exclusive SPI Monte Carlo simulated data and low-luminosity di-jet ATLAS data.
The final result for DPI event fraction from the best fit, multiplied by the pile-up correction factor, is
\begin{equation}
 {f}_{\rm DP} = 0.076 \pm 0.013 ({\rm stat.}) \pm 0.018 ({\rm syst.}).
 \label{myska:eq:f_DP}
\end{equation} 
\noindent The systematic uncertainty accommodates statistical uncertainty of the pile-up correction factor, theoretical uncertainty
obtained by alternative modeling of template A using Sherpa v1.3.1 \cite{Gleisberg:2009:jhep0902_007}
and by varying of the $p_{\rm T}^{\rm p}$ cut, as well as experimental uncertainties in the jet energy scale and resolution, 
physics background modeling and lepton response.

The evaluation of $\sigma_{\rm eff}$ is performed by inserting (\ref{myska:eq:f_DP}) into (\ref{myska:eq:sigma_eff_eval})
and the final result is
\begin{equation}
\sigma_{\rm eff} = 15 \pm 3 ({\rm stat.})^{+5}_{-3} ({\rm syst.}) ~{\rm mb}.
\end{equation} 
\noindent Other sources of the uncertainty, besides the $f_{\rm DP}$, are the physics background modeling, lepton response and measurement of the luminosity.

\section{Summary}

Using its entire 2010 data, ATLAS has measured the scaling factor for double-parton scattering.
The central value of the searched $\sigma_{\rm eff}$ was found to be 15 mb, which is consistent with the previous measurements.
The analysis was based on Monte Carlo simulation of exclusive single-parton interaction
producing $W$ boson in direct association with exactly two jets. It was found that 
these events create around 92$\%$ of the ATLAS data, after subtraction of physics background,
while the remaining events were identified with the double-parton scattering.

\end{document}